\documentclass{vgtc}                          

\ifpdf
  \pdfoutput=1\relax                   
  \usepackage{graphicx}                
  \DeclareGraphicsExtensions{.pdf,.png,.jpg,.jpeg} 
\else
  \usepackage{graphicx}                
  \DeclareGraphicsExtensions{.eps}     
\fi%

\usepackage{comment}
\usepackage{xspace}

\graphicspath{{figures/}{pictures/}{images/}{./}} 

\usepackage{microtype}                 
\PassOptionsToPackage{warn}{textcomp}  
\usepackage{textcomp}                  
\usepackage{mathptmx}                  
\usepackage{times}                     
\usepackage{cite}                      
\usepackage{tabu}                      
\usepackage{booktabs}   
\usepackage{amssymb}
\usepackage{enumitem}

    \newcommand{\PUR}[1]{\textcolor{black}{#1}}

\newcommand{\fpattern}{$PATTERN$\xspace}
\newcommand{\fdensity}{$DENSITY$\xspace}
\newcommand{\foutline}{$OUTLINE$\xspace}
\newcommand{\fbackground}{$BKG$\xspace}
\newcommand{\task}{$TASK$\xspace}
\newcommand{\trial}{$TRIAL$\xspace}

\newcommand{\tfaint}{$T_{Faintest}$\xspace}
\newcommand{\tstrong}{$T_{Strongest}$\xspace}

\newcommand{\pfill}{$P_{Fill}$\xspace}
\newcommand{\pdots}{$P_{Dots}$\xspace}
\newcommand{\pstripes}{$P_{Stripes}$\xspace}

\newcommand{\dlow}{$D_{Low}$\xspace}
\newcommand{\dhigh}{$D_{High}$\xspace}

\newcommand{\btrees}{$B_{Trees}$\xspace}
\newcommand{\bnull}{$B_{\varnothing}$\xspace}

\newcommand{\outlineon}{$O_{On}$\xspace}
\newcommand{\outlineoff}{$O_{\varnothing}$\xspace}

\newcommand{\hdensityfaint}{$\mathbf{H_{DensityFaint}}$\xspace}
\newcommand{\hdensitystrong}{$\mathbf{H_{DensityStrong}}$\xspace}
\newcommand{\hbackground}{$\mathbf{H_{Background}}$\xspace}
\newcommand{\houtline}{$\mathbf{H_{Outline}}$\xspace}

\onlineid{1136}

\vgtccategory{Research}

\vgtcinsertpkg

\title{A Study of Opacity Ranges for Transparent Overlays in 3D Landscapes}

\author{Jan Hombeck\thanks{e-mail: jan.hombeck@uni-jena.de}\\ %
        \parbox{1.4in}{\scriptsize \centering University of Koblenz \\ University of Victoria}\\
\and Li Ji\thanks{e-mail: li.ji@llamazoo.com}\\ %
     \scriptsize LlamaZOO Interactive Inc. %
\and Kai Lawonn\thanks{e-mail: kai.lawonn@uni-jena.de}\\ %
     \scriptsize University of Jena %
\and Charles Perin\thanks{e-mail: cperin@uvic.ca}\\ %
     \scriptsize University of Victoria}

\teaser{
  \centering
  \includegraphics[trim=0 10 0 20, clip,width=0.9\linewidth]{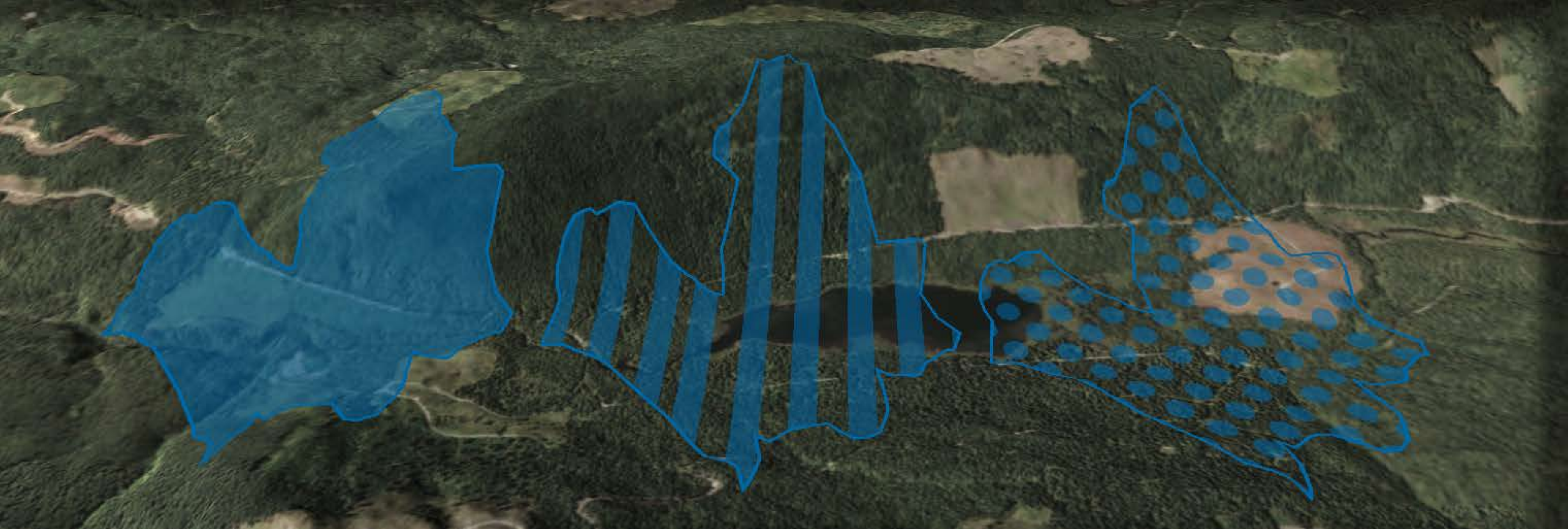}
\caption{Transparent overlays in a photorealistic 3D environment using striped, filled and dotted internal patterns with an outline.}
	\label{fig:teaser}
}

\abstract{When visualizing data in a realistically rendered 3D virtual environment, it is often important to represent not only the 3D scene but also overlaid information about additional, abstract data. These overlays must be usefully visible, i.e. be readable enough to convey the information they represent, but remain unobtrusive to avoid cluttering the view. We take a step toward establishing guidelines for designing such overlays by studying the relationship between three different patterns (filled, striped and dotted patterns), two pattern densities, the presence or not of a solid outline, two types of background (blank and with trees), and the opacity of the overlay. For each combination of factors, participants set the faintest and the strongest acceptable opacity values. Results from this first study suggest that i) ranges of acceptable opacities are around 20-70\%, that ii) ranges can be extended by 5\% by using an outline, and that iii) ranges shift based on features like pattern and density. 

} 

\CCScatlist{
  \CCScatTwelve{Overlays} {Opacity} {Visualization}{}
}

\begin{document}


\firstsection{Introduction and Background}

\maketitle

We investigate which opacity ranges to use when overlaying shapes for visualizing abstract data within a spatially bounded region of interest (RoI) on a photorealistic landscape. 
Our research is motivated by the recent adoption of real-time and high-performance photorealistic rendering in industrial data visualization applications, and in particular, high-fidelity GIS-based visualization applications~\cite{digitaltwin,cesiumstory,ESRIArcGIS}. 
The standard way of representing information within a RoI is to overlay a transparent layer of color on top of the realistically rendered landscape to delineate a region of interest~\cite{GeoVisBook}. This technique is used for example in map-based Geographic Information System (GIS) tools such as ESRI ArcGIS~\cite{ESRIArcGIS}. 
Creating an overlaid visualization that sits on top of a photorealistic, 3D environment, such as the example shown in Figure~\ref{fig:teaser} requires a visualization designer to make choices. 
These choices are currently limited, as they are inherited from 2D GIS tools and constrained to a transparent layer of color being displayed on top of the realistically rendered landscape~\cite{GeoVisBook}.
To think beyond this conventional design we can turn to the field of information visualization, which provides many empirical guidelines for representing abstract data, established through graphical perception studies~\cite{cleveland1984graphical}. 
Opacity is an effective visual variable to encode quantitative information~\cite{munzner-book} such as uncertainty~\cite{collins2007visualization} of the RoI, and has been the focus of studies on the perception of grids in 2D charts~\cite{bartram-whisper}.
Graphical Perception studies in information visualization, however, make strong assumptions regarding the background and environment in which the abstract data is being represented, typically a uniform background or a 2D map and a static point of view. 
In contrast, 3D landscapes are colorful, contain complex geometries and textures, and can be looked at from infinite numbers of points of view. 
It makes it difficult to directly apply results from studies that consider standard 2D charts (e.g.,~\cite{gutwin-emphasis,haase2000meteorology,hughes2001just,bartram-whisper}).



\section{Study of Opacity Ranges for Overlays}
\label{sec:study}

To identify reasonable opacity ranges for overlays, we must consider the many other graphical properties of these overlays. Our first step was to identify promising ones to study. 

\PUR{In the field of illustrative rendering, patterns that are filled~\cite{gooch}, striped~\cite{ware2004view,interrante1996illustrating} and doted~\cite{martin2017survey,interrante1996illustrating} are widely used to render overlays.}

\PUR{Outlines \cite{lawonn2016feature} are commonly used to increase the visibility of 3D shapes~\cite{lawonn2018survey,lawonn2018surveyMed}. 
Pattern density~\cite{durgin1996visual} and overlay opacity~\cite{bartram-whisper} also affect how information is perceived.}
\PUR{With transparent overlays, the rendered image must contain sufficient visual cue to permit the Human Visual System (HVS) to recognize the separate layers, instead of a single layer of the blended colors~\cite{metelli1974perception, d1997color}. }

A sketching session with our research group and an interview with a visualization practitioner confirmed the usefulness of investigating the following rendering parameters, widely used in the literature: pattern style, pattern density, pattern outline, and scene background.



\subsection{Apparatus}
We conducted a remote study in accordance to our research ethics board regulations during the covid-19 pandemic.
A remote study means lower \textbf{internal validity} of the results~\cite{mcgrath1995methodology} because the apparatus is less controlled than in a lab setting (e.g., varying display sizes, luminosity and contrast).
To control for these factors we required participants to use a monitor with resolution of 1920x1080 px and with the luminosity value of their screen set to its maximum, and we conducted the study during daylight hours only. 
On the other hand, a remote study increases the \textbf{external validity} of the results~\cite{mcgrath1995methodology}. It is closer to the real-world context, where people use various devices and screens in various environments.
Overall, our study loses some precision to increase its realism and generalizability~\cite{mcgrath1995methodology}.

We conducted the study with the TeamViewer remote desktop control software~\cite{TeamViewerSoftware}. 
This allowed the experimenter to verbally communicate with the participants, answer their questions and observe their interactions. 
Participants interacted with the study software via mouse and keyboard.
While we required a good internet connection from participants, the delay (time between the action from the participant and the response from the server) varied between participants and sometimes within a session.
The study software showed views of a forest area in a professionally used virtual environment in Unity.

\subsection{Tasks}

To establish opacity ranges in which different configurations of overlay are considered useful, we created the two following tasks based on Bartram et al.'s\cite{bartram-whisper} study of opacity values for grids:

\begin{itemize}[noitemsep,nolistsep]
\setlength\itemsep{0em}
    \item \textit{\textbf{\tfaint}: Please adjust the color to be as faint as you think it can comfortably be to be still useful; any fainter and you would no longer be able to easily use it.}
    \item \textit{\textbf{\tstrong}: Please adjust the color to be as strongly visible as you think it can comfortably be before it interferes with or ``comes in front of'' the environment; any stronger and it would be too obtrusive.}
\end{itemize} 

Participant used the left and right arrow keys to change the opacity value and pressed the space bar to confirm and go to the next trial.


\begin{figure}[t]
    \centering
    \includegraphics[width=\linewidth]{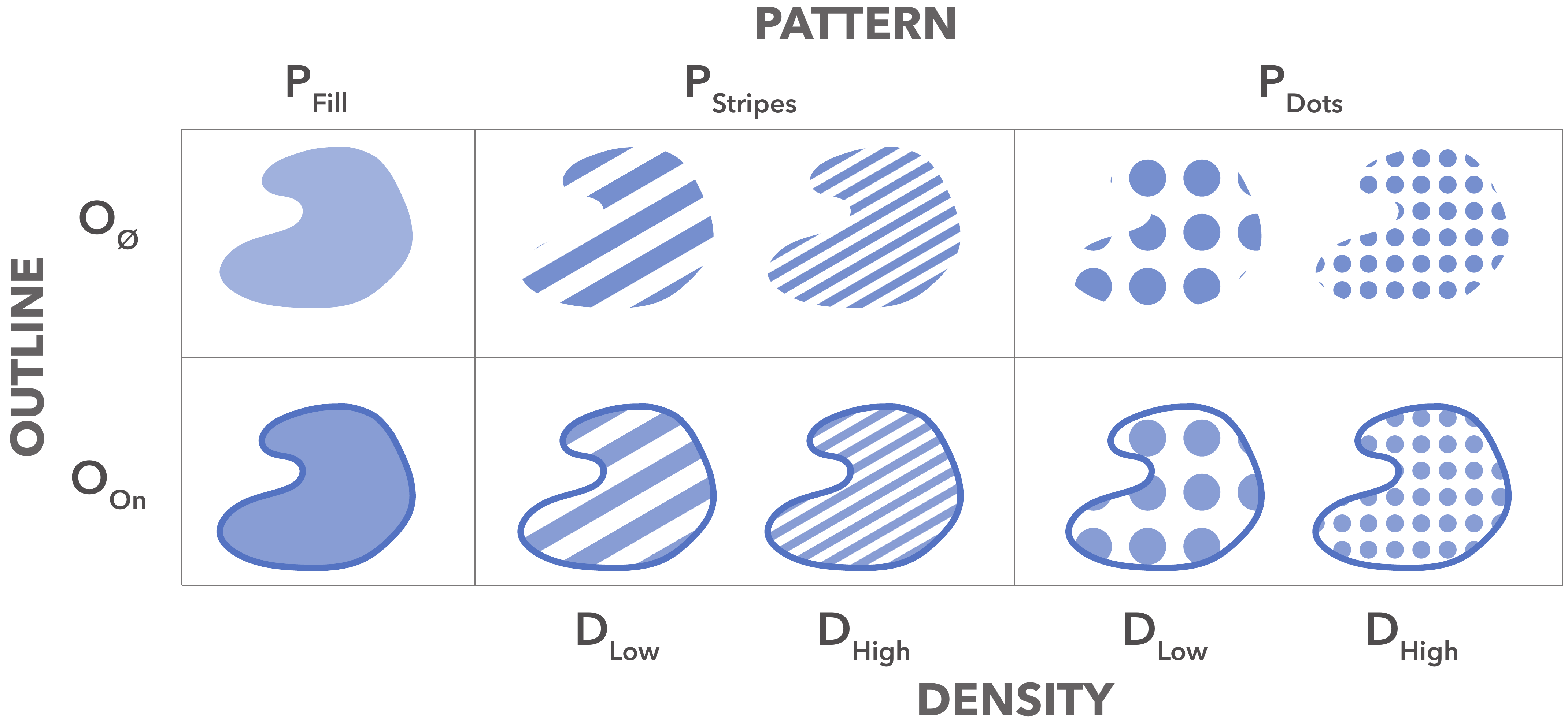}
    \vspace{-1em}
    \caption{The three \fpattern for the two \foutline for the two \fdensity used in the study (only \fbackground is not shown in this Figure).}
    \label{fig:factors}
\end{figure}


\subsection{Factors and Experimental Design}
\label{sec:exp-design}
\begin{figure}[t]
    \centering
    \includegraphics[trim=0 180 0 180, clip,width=\linewidth]{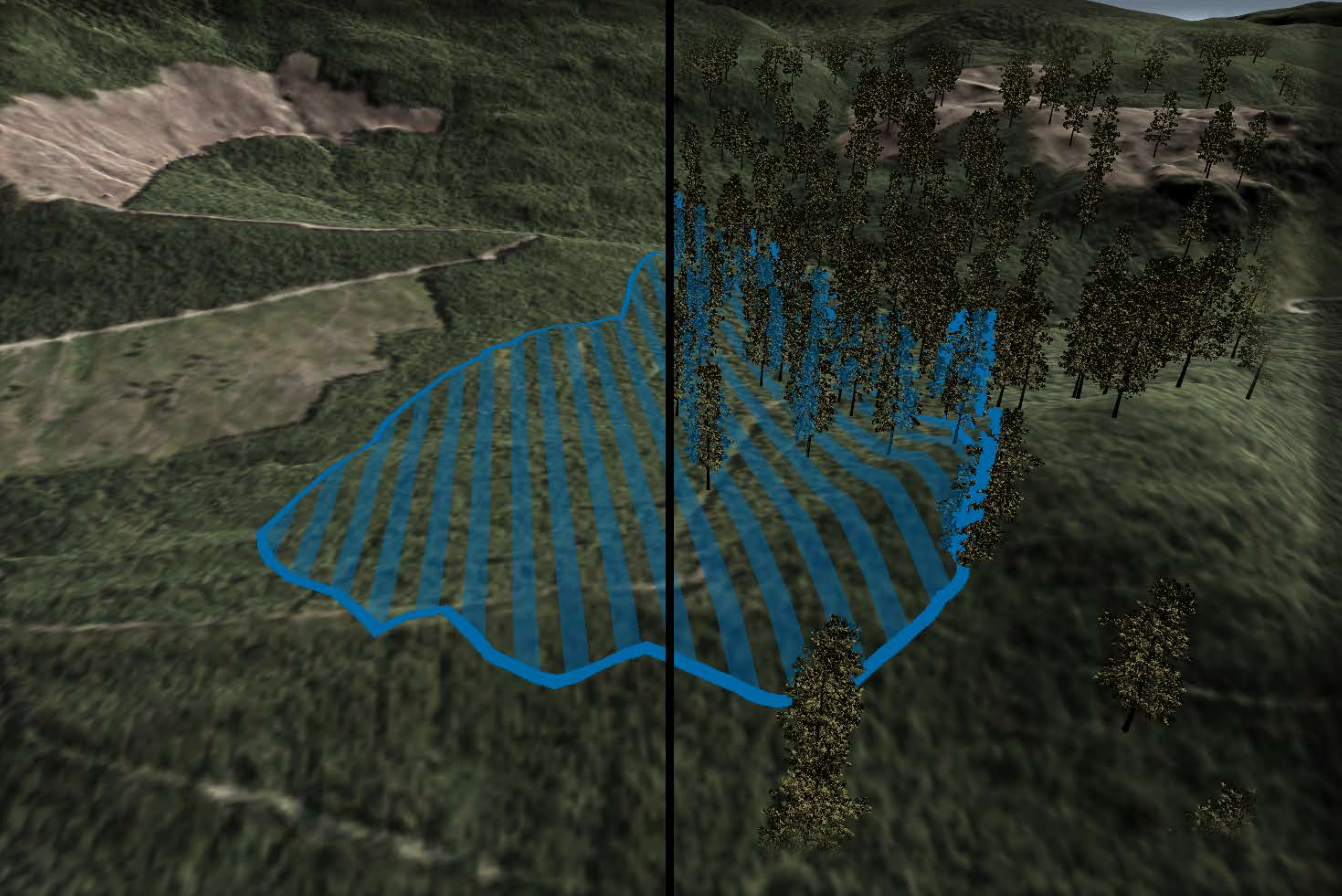}
    \vspace{-1em}
    \caption{Side by side view of two configurations used during the study. \pstripes, \dhigh, \outlineon with \bnull (left) and \btrees (right). 
    \PUR{The overlay is rendered in a post-processing step that maps its shape on top of the underlying surface (that includes objects present in the scene).} }
    \label{fig:factor-background}
\end{figure}

We studied the four following factors (see Figures~\ref{fig:factors} and~\ref{fig:factor-background}):
\begin{itemize}[noitemsep,nolistsep]

    \item \fpattern was either \pfill (a solid color pattern), \pstripes (a stripped pattern). or \pdots (a dotted pattern). 
    
    \item Both \pstripes and \pdots were presented with two different values for \fdensity. 
    With \pstripes at low density \dlow, each stripe as a 240-pixel width and a 120-pixel width at high density \dhigh.
    With \pdots each dot has a 240-pixel diameter at \dlow and a 120-pixel diameter at \dhigh. 
    
    \item The overlay \foutline (a 12-pixel width solid stroke around the RoI) was either present (\outlineon) or not (\outlineoff). 
    
    \item The scene \fbackground was either with trees \btrees or without \bnull. 

\end{itemize}

\noindent
Overall the experiment consisted of 
$[ 2 \times$\fpattern (\pstripes, \pdots) $\times 2 \times$\fdensity (\dlow, \dhigh)
$+ 1 \times$\fpattern (\pfill)$]$
$\times 2 \times$\foutline (\outlineon, \outlineoff)
$\times 2 \times$\fbackground (\bnull, \btrees)
$\times 2 \times$\task (\tfaint, \tstrong)
$\times 4 \times$\trial = 160 TRIALS per participant.
Each participant performed the experiment in two task blocks, one for \tfaint and one for \tstrong (80 trials per task block in random order). The order of task block was balanced across participants.

\begin{figure*}[t]
    \centering
    \includegraphics[width=\linewidth]{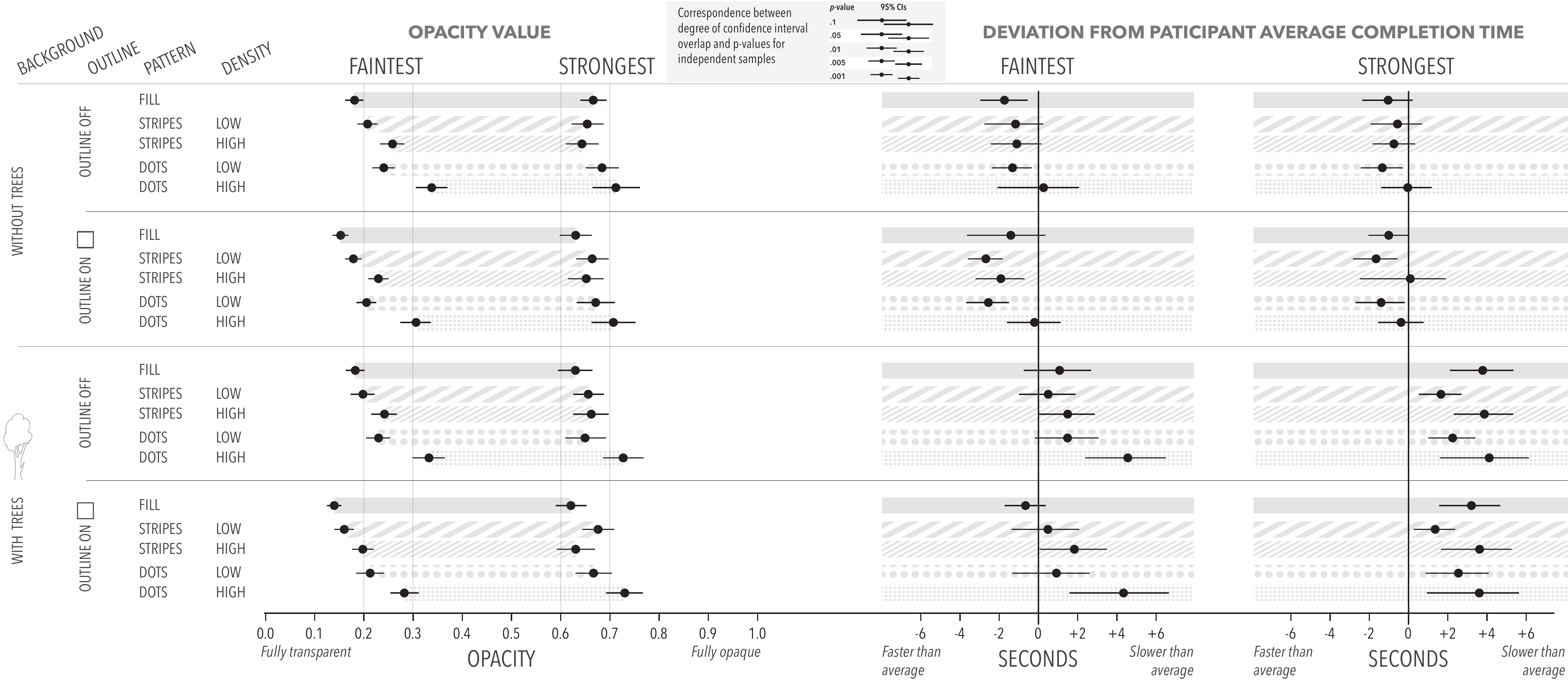}
    \caption{Opacity values and deviation from average completion time mean 95\% bootstrappd CIs for each task, grouped by \fbackground, then \foutline, then \fpattern and then \fdensity.
    The correspondence between degree of CI overlap and p-values for independent samples is based on~\cite{krzywinski2013points}. 
    The rectangles between mean estimates for opacity values give an indication of the width of the opacity range for each condition. 
    }
    \label{fig:results-ci-all}
\end{figure*}

\subsection{Procedure}
\label{sec:procedure}



The experimenter introduced the study and let the participant read the instructions explaining the tasks.
The participant then performed a series of training trials until they understood the tasks and controls. 
After having asked any question they had, they started the recorded trials, either with \tfaint or with \tstrong first.
Once they had completed the first task block, they were asked to take a 5-minute break before completing the second task block.
Finally they filled out the study questionnaire and participated in a short debriefing. 
The questionnaire asked demographic questions (age, gender, level of education, experience in visualization, experience in 3D, usage of video games and of computers) and study-related questions on Likert scales. Each session lasted approximately one hour.


\subsection{Hypotheses}
\PUR{We hypothesize that the denser a pattern, the easier to perceive the overlay as a conceptually separate layer for the HSV, and thus the less saliency the overlay will need.
For instance, \tfaint will lead to higher opacity values for \dlow, because a low-density pattern will require more emphasis through higher opacity, for the HVS to recognize the overlay as a conceptually separate layer.}

\PUR{We also hypothesize that the outline will facilitate the HVS to more easily recognize that the overlay belongs to a separate shape, and therefore belongs to a conceptually separate transparent layer.
Therefore, the outline will make a wider range of opacity recognizable as transparent overlays instead of a single layer of blended color.  }

More formally, our hypotheses were as follows:
\begin{itemize}[noitemsep,nolistsep]
    \item \hdensityfaint: \dlow will lead to higher average opacity values than \dhigh for \tfaint. 
    Considering that \pfill has the highest density, we therefore hypothesize that for \tfaint, 
    O(\pfill) $<$ O(\pstripes $\times$ \dhigh) $<$ O(\pdots $\times$ \dhigh) $<$ O(\pstripes $\times$ \dlow) $<$ O(\pdots $\times$ \dlow).
    Indeed, with \dlow, the pattern shapes are larger and there are fewer of them, resulting in relatively large empty spaces in the pattern. 
    
    \item \hdensitystrong: \dlow will lead to higher average opacity values than \dhigh for \tstrong. 
    With \pfill having infinite density, we hypothesize that for \tstrong, we will obtain
    O(\pfill) $<$ O(\pstripes $\times$ \dhigh) $<$ O(\pdots $\times$ \dhigh) $<$ O(\pstripes $\times$ \dlow) $<$ O(\pdots $\times$ \dlow).
    Indeed, we hypothesize that larger empty spaces make it easier to mentally reconstruct the underlying 3D scene thus tolerate greater opacity values.
    
    \item \houtline: Outlines will enlarge opacity ranges, i.e. for \tfaint, O(\outlineon) $<$ O(\outlineoff) and for \tstrong, O(\outlineoff) $<$ O(\outlineon), because the shape of the overlay is clearer with an outline. 
    
    \item \hbackground: Trees in the background will lead to smaller opacity ranges, i.e. for \tfaint, O(\bnull) $<$ O(\btrees), because additional content such as trees distort the overlay; and for \tstrong, O(\btrees) $<$ O(\bnull), because comfortably seeing the trees might require lower opacity values.
    

\end{itemize}

\subsection{Participants}
We recruited sixteen participants (six male, ten female) with normal or corrected vision through mailing lists and social networks. 
Most participants were between 19--25 years of age with on average 1-2 years of experience with 3D environments and data visualization. 
Detailed participant information is provided in supplemental material.
Participants received \$15 via email for their participation.

\subsection{Results}


We base our analyses on \emph{estimation} using 
bootstrapped confidence 
intervals instead (CI) of p-values, following recommendations from 
APA~\cite{apa:2010:publication_manual}.
A 95\% CI contains the true mean 95\% of the 
time and conveys effect sizes~\cite{cumming:2005:inference_by_eye}. 
We pre-specified all analyses before running the study and
tested on pilot data (scripts for parsing the
data, computing CIs, and generating drafts of figures).


\subsubsection{Confirmatory Analysis}
Here we revisit our hypotheses against the results shown in Figure~\ref{fig:results-ci-all}. 

\hdensityfaint is partially confirmed. Indeed, \pfill has consistently lower opacity values than the other configurations of \fpattern and \fdensity. However, \pstripes $\times$ \dlow has a significantly lower opacity value than  \pstripes $\times$ \dhigh; and \pdots $\times$ \dlow has a significantly lower opacity value than  \pdots $\times$ \dhigh -- these results are opposite to what we hypothesized.
\hdensitystrong is not confirmed, with only \pfill and \pstripes $\times$ \dhigh leading weakly to greater opacity values than \pdots $\times$ \dhigh (around 5\% greater).
Overall, 
lower density leads to lower values for the lower bound of the opacity range, 
resulting in larger usable opacity ranges (around 5\% larger for \pstripes and 10\% larger for \pdots).


\houtline is not confirmed for \tfaint with \bnull and \tstrong, with no notable difference based on \foutline. 
However, there are differences for \tfaint and with \btrees, for which \pfill, \pstripes $\times$ \dlow, \pstripes $\times$ \dhigh and \pdots $\times$ \dlow have smaller opacity values (around 5\% smaller) with \outlineon than with \outlineoff, i.e. the opacity ranges are extended by around 5\% (except for \pdots $\times$ \dhigh).

\hbackground is not confirmed as there is no strong difference between any of the conditions based on \fbackground.

\subsubsection{Exploratory Analysis}

The results reveal additional aspects beyond hypotheses checking.
First, different conditions lead to different opacity ranges. 
Opacity ranges for \pfill are consistently slightly larger than for \pstripes $\times$ \dlow, which are larger than \pdots $\times$ \dlow, which are larger than \pstripes $\times$ \dhigh, which are larger than \pdots $\times$ \dhigh.

Second, the CIs for \tstrong are wider than for \tfaint, i.e. there is higher variance and uncertainty with \tstrong than with \tfaint. 

Third, Figure \ref{fig:results-ci-all} shows the deviation from participant average completion time for both tasks (instead of absolute completion times that are meaningless with the remote setup).
A configuration led to faster completion if its CI falls on the left side of the 0 baseline, and to slower completion if it falls on the right side.
We notice that \fbackground strongly affects participant completion time, with nearly all conditions with \btrees resulting in longer completion times than average. 
We also observe that \pfill and \pstripes $\times$ \dlow might result in slightly shorter times than \pdots $\times$ \dhigh.



\subsubsection{Questionnaire Answers}
14/16 participants selected \pfill as their favorite pattern.
\pfill also received the strongest scores for performing both \tfaint (12 participants strongly positive) and \tstrong (13 strongly positive), and for confidence in their answers (14 strongly positive). 
\pstripes received mitigated scores for \tfaint (7 positive, 6 negative) but was better ranked for \tstrong (10 positive, 3 negative), and made participants quite confident in their answers (12 positive, 2 negative).
\pdots was the least preferred pattern, with 7 positive and 6 negative for \tfaint, 7 positive and 6 negative for \tstrong, and 6 positive and 5 negative for confidence in their answers.
Participants found that \outlineon helped for completing both \tfaint (13 positive) and \tstrong (11 positive), and increased their confidence in their answers (13 positive).

\section{Discussion}
\label{sec:discussion}

\subsection{What is Wrong with Density?}
We did not expect that increasing the density of a pattern would also increase the required opacity value, with the effect being stronger for \tfaint. 
It is possible that a more complex pattern makes it more challenging to mentally reconstruct the information about two layers (the overlay and the 3D scene underneath) from a single image, thus requires stronger visual stimuli and salience, i.e. higher overlay opacity value.
Completion times conform with this observation. 
The time required to set the opacity for \dhigh is in almost all cases longer than for \dlow, which may be due to longer times spent to mentally reconstruct the overlay. 
Since for \tstrong the pattern is more visible the effect is not as strong as for \tfaint. 


\subsection{Different Patterns, Different Opacity Ranges}
\label{sec:pattern}

Some overlay configurations have wider opacity ranges than others. 
\pdots has relatively smaller, shifted upwards ranges, compared to \pfill and \pstripes (i.e. higher minimum and maximum bounds). 
We think this is again due to the complexity of the pattern,
and that higher opacity values for the overlays with \pdots make the distinction of the two layers (overlay and 3D scene underneath) easier. 

\begin{figure}[t]
    \centering
    \includegraphics[width=1.0\linewidth]{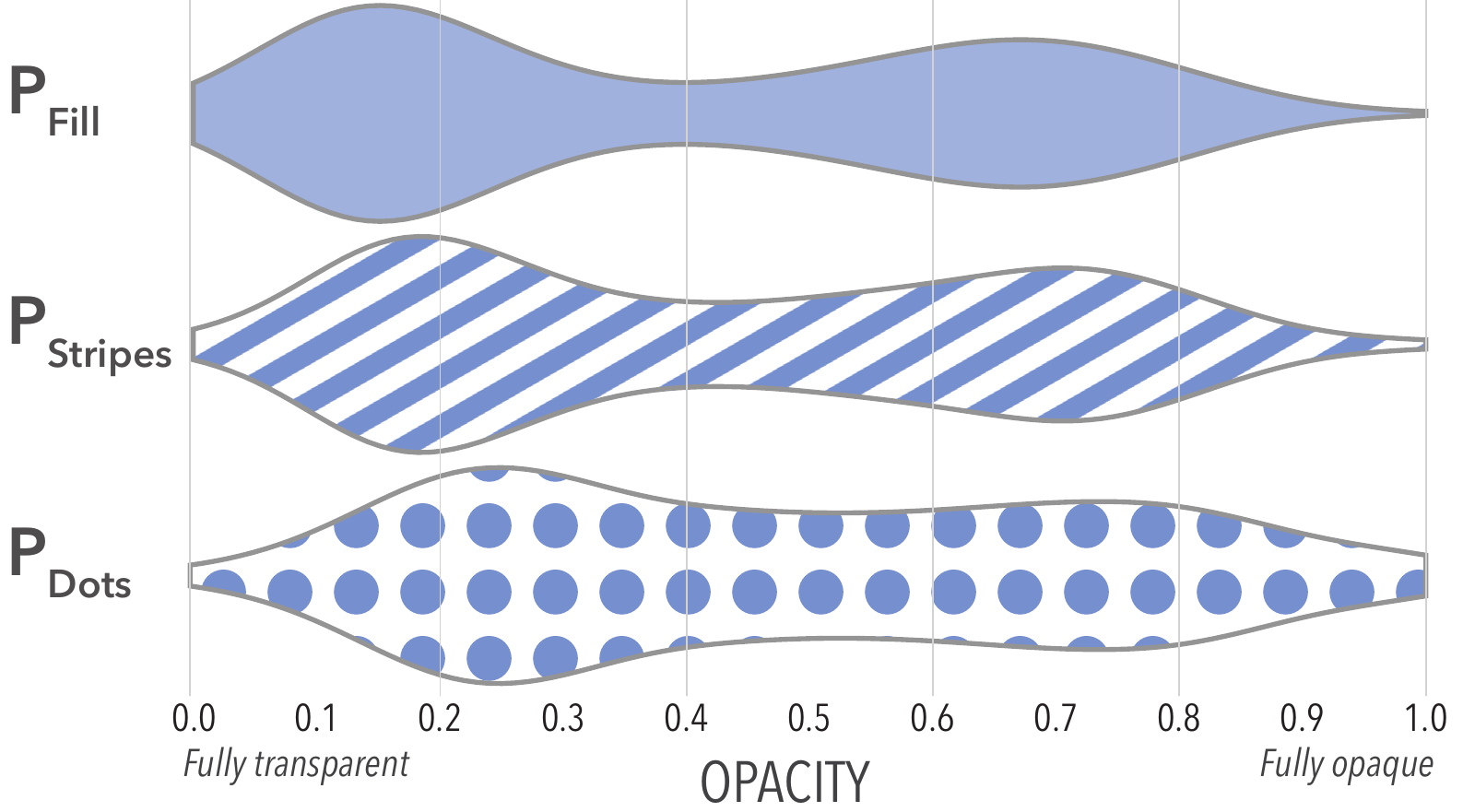}
        \vspace{-1em}
    \caption{Distributions of all opacity values for each \fpattern. 
    }
        \vspace{-0.5em}
    \label{fig:violinPattern}
\end{figure}

Figure~\ref{fig:violinPattern} shows differences in distributions of opacity values used for each \fpattern. 
All three plots have two inflexion points, for their minimum and maximum bound of opacity range, but their overall shapes differ. 
The \textbf{extent} of these opacity ranges, combined with their degree of \textbf{bimodality} (how separable the minimum bound is from the maximum bound) can guide the selection of rendering parameters according to the data to visualize. 
For example, an overlay configuration with a small extent and a low degree of bimodality would be appropriate to visualize a binary data dimension, e.g enable or disable the overlay;
a high degree of bimodality is appropriate for a quantitative data dimension; and a large extent is appropriate for a quantitative data dimension that requires a high resolution. 


\subsection{Terrain, Objects and Outlines}

Our results show that an outline can help increase the acceptable opacity range of an overlay, especially on the lower opacity bound. 
With an outline the pattern is no longer solely responsible for establishing the structure of the overlay, it serves as an additional internal support and can be less salient. 
Outlines are particularly useful when there are objects in the scene, increasing the range of acceptable opacity values (lower bound) by around 5\% in that case.
Outlines are less useful for \tstrong, because the region of interest is already sufficiently visible due to the salience of the pattern itself. 

We found that participants were slower with objects (trees) in the background, for all conditions we tested.
We explain this result because the rendering of an overlay is partially occluded by the objects inside. 
It results in a more complex overlay shape with uncertain boundaries that takes more time to reconstruct mentally.

Several participants found that \pstripes helped identify the structure of the terrain. 
However, we noticed that this happened mostly when the stripes were parallel to the slope of the terrain.
One way to improve structural visibility would be to use grid-like patterns that would  assist in understanding the gradient in two dimensions. 

\subsection{Looking Forward}
Our empirical results provide a starting point to establish how to design overlaid visualizations of abstract data in realistically rendered 3D environments.
This first step necessarily comes with limitations in terms of generalizability of the results.

\PUR{First, the study measures people's preferences for opacity ranges instead of visual performance. This is a subjective measure that allows us to derive the most comfortable opacity ranges for the user. In the future we plan to conduct quantitative perceptual studies that will help us measure more objectively visual performance.}

Second, we used a landscape with and without trees, but other aspects of the terrain need to be investigated: what if the texture of the terrain is light, like snow, or dark, like grass or forest? What if it is uniform, like an ocean, or has many topological features, like a mountain?
There are also many more possible design variations of the overlays to study.
These include the color of the overlay, the contrast between the dominant color of the terrain and the color of the overlay, the orientation of the patterns, the distribution of points and lines in the patterns and the point shape and radius.

The world of computer graphics offers a wide array of additional rendering parameters that might not be traditionally used when visualizing abstract data on 2D charts (or even avoided by fear of creating chart junk), but that could be relevant for overlays in 3D environments. 
This includes the impact of color, contrast, shading or reflections.
These additional considerations may be valuable to make it easier to discriminate the overlay from the background when they are of similar color. In our study we kept the rendering as simple as possible to avoid artifacts or distractions, but one could for example look at casting shadows and incorporating reflections. 

Another important aspect to study is that of intersecting overlays. Having multiple (overlapping) RoI to analyze simultaneously is common within industrial applications. How to display multiple data sets is a current challenge that also calls for additional research. 


\section{Conclusion}
\label{sec:conclusion}
We have provided the first set of empirically determined acceptable (subjective) opacity ranges to use when adding overlays on top of a photorealistic 3D environment.
Through our study we observed participants set the lower and upper limits for opacity values for three different patterns (filled, striped, and dotted patterns), two pattern densities, the presence or not of a solid outline, and two types of background (blank and with trees).

We found that the range of acceptable opacity values is roughly between 20-70\%, with ranges shifting up or down slightly according to the overlay configurations.
While patterns like dots are mostly considered as distracting, we found that striped patterns can help better understand the curvature of the underlying terrain.

How to coalesce data visualization with realistic and interactive rendering in a unified visual context is an open research question with practical industrial impacts. 
\PUR{Rather than providing definite answers to this open question, our initial inquiry provides many directions worth studying to better understand how to represent overlays on top of complex, realistic 3D scenes. We look forward to continuing this exploration at the junction of information and scientific visualization, and realistic computer graphics rendering. }

\acknowledgments{\PUR{
This work was funded by NSERC, the Carl Zeiss Foundation and the Federal Ministry for Economic Affairs and Energy of Germany.}}

\bibliographystyle{abbrv-doi}

\bibliography{template}
\end{document}